\documentstyle[aps]{revtex}
\begin{document}
\bibliographystyle{prsty}
\baselineskip=8.5mm
\parindent=7mm
\begin{center}
{\large {\bf \sc{Nonfactorizable soft contributions in the $B\longrightarrow \eta_c K,\chi_{c0} K$   decays with the light-cone sum rules approach}}} \\[2mm]
Zhi-Gang Wang$^{1,2,3}$ \footnote{E-mail,wangzgyiti@yahoo.com.cn;  zgwang1973@126.com .}, Lin Li$^{2}$ and Tao Huang$^{1,2}$    \\
$^{1}$ CCAST (World Laboratory), P.O.Box 8730, Beijing 100080,
P. R. China \\
$^{2}$ Institute of High Energy Physics, P.O.Box 918 ,Beijing 100039,P. R. China \\
$^{3}$ Department of Physics, North China Electric Power University, Baoding 071003, P. R. China \footnote{Mailing address}\\

\end{center}

\begin{abstract}
In this article, we calculate the nonfactorizable soft
contributions in the $B \rightarrow \eta_c K, \chi_{c0} K$ decays
with the light-cone QCD sum rules approach. Our results show that
the nonfactorizable corrections from the soft gluon exchanges in
the decay $B \rightarrow \eta_c K$ are of $(15-30)\%$ and should
 be taken into account. As for the decay $B \rightarrow \chi_{c0} K$,
the factorizable contributions are zero and the nonfactorizable
 contributions from the  soft hadronic matrix elements  are
too small to accommodate the experimental data.
\end{abstract}


\section{Introduction}
The nonleptonic decays of  the $B$ meson have attracted much
attention in studying the
 nonperturbative dynamics of QCD, final state interactions and CP violation.
The exclusive  $B$  to  charmonia decays  are important since
those decays $B \to J/\psi K, \eta_c K, \chi_{cJ} K $ are regarded
as the golden channels for the study of CP violation. The
quantitatively   understanding of those decays depends  on our
knowledge about the nonperturbative hadronic matrix elements of
the operators entering the effective weak Hamiltonian. In
Ref.\cite{BBNS}, the authors propose an original approach called
 QCD-improved factorization  to deal with the
two-body nonleptonic  decays of the B meson. In this approach, the
decay amplitudes are expressed in terms of the semileptonic form
factors, hadronic light-cone distribution amplitudes and
hard-scattering  amplitudes. The semileptonic form factors, the
light-cone distribution  amplitudes are taken as input parameters
and the hard-scattering amplitudes including nonfactorizable
corrections due to the exchanges of hard gluons are calculated by
perturbative QCD.  For the exclusive $B$  to   charmonia decays ,
the QCD-improved factorization approach is broken  down due to the
divergence arising from the soft-gluon exchanges, moreover, the
theoretical branching  fractions are too small to accommodate the
experimental data \cite{BBNS,ChKT,Chao2,chay,cheng}. In Refs.
\cite{ChKT,Chao2}, the authors observe that for the exclusive
$B\rightarrow \eta_c K$ decay,
 the nonfactorizable
corrections to the naive factorization are infrared safe at
leading-twist order, the spectator interactions arising from the
kaon twist-3 effects are formally power-suppressed but chirally
and logarithmically enhanced; for the $B\rightarrow \chi_{c0} K$
decay, there are infrared divergences arising from the
nonfactorizable vertex corrections as well as logarithmic
divergences due to the spectator interactions even at
leading-twist order.

The effects of soft gluons which break down  factorization are
supposed of order $O(\Lambda_{QCD}/m_b)$ and neglected in the
QCD-improved factorization studies, however, no theoretical work
has ever proved that they are small quantities. For the
color-suppressed $B$ to charmonia decays, there may be significant
impacts of the nonfactorizable soft contributions.

On the other hand, the QCD light-cone sum rules (LCSR) approach
provides a powerful tool for calculating the exclusive soft
hadronic amplitudes \cite{Balitsky,Braun1,Chernyak1}. The LCSR
approach carries out the
 operator product expansion   near the light-cone $x^2\approx 0$
 instead of the short distance $x\approx 0$ while the nonperturbative
 matrix elements  are parameterized by the light-cone distribution  amplitudes
 which classified according to their twists  instead of
 the vacuum condensates. For detailed discussions of this approach,
 one can consult Ref. \cite{Kho1}. The LCSR approach has been applied to study the nonfactorizable
hadronic matrix elements due to the soft gluons exchanges and gave satisfactory results \cite{Kh2001,Melic,Tao}.

Experimentally, the $B \rightarrow \eta_c K$ decay was observed by
CLEO, BaBar, Belle Collaborations and the $B \rightarrow \chi_{c0} K$ decay
 by Belle Collaboration with relatively large branching fractions \cite{cbb,Abe}. The
large discrepancies between the theoretical and experimental
values for those decays call for considerations  of new
ingredients and mechanisms. It is interesting to study the
nonfactorizable soft contributions in the
 pseudoscalar and scalar charmonia decays $B \rightarrow \eta_c K,\chi_{c0}K$ with the LCSR approach.

The article is organized as follows:   the factorizable
contributions from the effective weak Hamiltonian are derived
in Sec.II; the soft hadronic matrix elements $\langle \eta_c K|
{\cal \widetilde{O}} | B \rangle $ and $\langle \chi_{c0} K| {\cal
\widetilde{O}} | B \rangle $ are calculated with the light-cone
sum rules  approach in Sec.III;
 numerical results  are presented in Sec.IV;
the section V is reserved for conclusion.

\section{Effective weak Hamiltonian and factorizable contributions}

The effective weak Hamiltonian for the $b\rightarrow s c c$ decay
modes can be written as (for detailed discussion of the effective
weak Hamiltonian, one can consult Ref. \cite{Buras})
\begin{equation}
H_w = \frac{G_F}{\sqrt{2}}\left\{V_{c b} V_{c s}^*
\left[ C_1(\mu) {\cal O}_1 +  C_2(\mu) {\cal O}_2 \right] -V_{tb} V_{ts}^*\sum_{i=3}^{10}
C_i {\cal O}_i \right\}  \, ,
\end{equation}
where $V_{ij}$'s are the CKM matrix elements, $C_i$'s are the
Wilson coefficients calculated at the renormalization scale $\mu
\sim O(m_b)$ and the relevant operators ${\cal O}_i$  are given by
 \begin{eqnarray}
&& {\cal O}_1=(\overline{s}_{\alpha} b_{\beta})_{V-A}
(\overline{c}_{\beta} c_{\alpha})_{V-A},\qquad\qquad~ {\cal
O}_2=(\overline{s}_{\alpha} b_{\alpha})_{V-A}
(\overline{c}_{\beta} c_{\beta})_{V-A},
 \nonumber\\
&& {\cal O}_{3(5)}=(\overline{s}_{\alpha} b_{\alpha})_{V-A}
\sum_q (\overline{q}_{\beta} q_{\beta})_{V-A(V+A)},~ {\cal
O}_{4(6)}=(\overline{s}_{\alpha} b_{\beta})_{V-A}  \sum_q
(\overline{q}_{\beta} q_{\alpha})_{V-A(V+A)},\nonumber
\\
&& {\cal O}_{7(9)}={3\over 2}(\overline{s}_{\alpha}
b_{\alpha})_{V-A}  \sum_q e_q (\overline{q}_{\beta}
q_{\beta})_{V+A(V-A)},~ {\cal O}_{8(10)}={3\over
2}(\overline{s}_{\alpha} b_{\beta})_{V-A}  \sum_q e_q
(\overline{q}_{\beta} q_{\alpha})_{V+A(V-A)}.
 \end{eqnarray}
We can reorganize  the color-mismatched quark fields into color
singlet states by Fierz transformation, (for example, $ {\cal
{O}}_1 = \frac{1}{N_c} {\cal {O}}_2 + 2 {\cal \widetilde{O}}_2
 $, $N_c$  is the color number and taken as $3$.)  and express the
effective weak Hamiltonian $H_w$ in the following form,
\begin{eqnarray}
H_w = \frac{G_F}{\sqrt{2}} &&  \left\{  V_{c b} V_{c s}^*
\left[ \left ( C_2(\mu) + \frac{C_1(\mu)}{3} \right ) {\cal O}_2 +  2 C_1(\mu) {\cal \widetilde{O}}_2 \right]
 \right. \nonumber \\
&& \left.-V_{tb} V_{ts}^*
\left[ \left ( C_3(\mu) + \frac{C_4(\mu)}{3} \right ) {\cal O}_3 +  2 C_4(\mu) {\cal \widetilde{O}}_3 \right]
\right. \nonumber \\
&& \left. -V_{tb} V_{ts}^*
\left[ \left ( C_5(\mu) + \frac{C_6(\mu)}{3} \right ) {\cal O}_5 +  2 C_6(\mu) {\cal \widetilde{O}}_5 \right]
\right. \nonumber \\
&& \left. -V_{tb} V_{ts}^*
\left[ \left ( C_7(\mu) + \frac{C_8(\mu)}{3} \right ) {\cal O}_7 +  2 C_8(\mu) {\cal \widetilde{O}}_7 \right]
\right. \nonumber \\
&& \left. -V_{tb} V_{ts}^*
\left[ \left ( C_9(\mu) + \frac{C_{10}(\mu)}{3} \right ) {\cal O}_9 +  2 C_{10}(\mu) {\cal \widetilde{O}}_9 \right]
\right\},
\end{eqnarray}
where
\begin{eqnarray}
&& {\cal O}_{2 (3,9)} = (\overline{c} \gamma_{\mu}(1-\gamma_5) c)(\overline{s} \gamma^{\mu} (1-\gamma_5)b) \;\; ,\;
{\cal \widetilde{O}}_{2 (3,9)} =  (\overline{c} \gamma_{\mu}(1-\gamma_5) \frac{\lambda_a}{2} c)
(\overline{s} \gamma^{\mu}(1-\gamma_5) \frac{\lambda_a}{2} b) \, ,  \nonumber \\
 && {\cal O}_{5(7)} = (\overline{c} \gamma_{\mu}(1+\gamma_5) c)(\overline{s} \gamma^{\mu} (1-\gamma_5)b) \;\; ,\;
{\cal \widetilde{O}}_{5(7)} =  (\overline{c} \gamma_{\mu}(1+\gamma_5) \frac{\lambda_a}{2} c)
(\overline{s} \gamma^{\mu}(1-\gamma_5) \frac{\lambda_a}{2} b)  \, ,
\end{eqnarray}
 here $\lambda^a$'s are $SU(3)$ Gell-Mann matrices.

The factorizable  matrix elements of the operator ${\cal O}_i$ for the decay $B\rightarrow \eta_c K$ can be parameterized  as
\begin{eqnarray}
 \langle \eta_c(p) K(q) |  H_w | B(p+q) \rangle
=\frac{G_F}{\sqrt{2}}&& \left\{ \left[ V_{c b} V_{c s}^* \left ( C_2(\mu) + \frac{C_1(\mu)}{3} \right )
  -V_{t b} V_{t s}^* \left ( C_3(\mu)+C_9(\mu) + \frac{C_4(\mu)+C_{10}(\mu)}{3} \right ) \right] \right. \nonumber \\
&& \langle \eta_c(p) | \overline{c} \gamma_{\mu} (1-\gamma_5) c
| 0 \rangle \langle K(q)| \overline{s} \gamma^{\mu} (1-\gamma_5) b | B(p+q) \rangle -V_{t b} V_{t s}^*\left ( C_5(\mu)+C_7(\mu) \right. \nonumber \\
&&  \left.  \left. + \frac{C_6(\mu)+C_8(\mu)}{3} \right )
\langle \eta_c(p) | \overline{c} \gamma_{\mu} (1+\gamma_5) c
| 0 \rangle
\langle K(q)| \overline{s} \gamma^{\mu} (1-\gamma_5) b | B(p+q) \rangle \right\}\, , \nonumber \\
=\frac{G_F}{\sqrt{2}}&& \left\{   -V_{c b} V_{c s}^* \left ( C_2(\mu) + \frac{C_1(\mu)}{3} \right )  +V_{t b} V_{t s}^* \left ( C_3(\mu)-C_5(\mu)-C_7(\mu)+C_9(\mu) \right.\right.\nonumber\\
&&\left.\left.+ \frac{C_4(\mu)-C_6(\mu)-C_8(\mu)+C_{10}(\mu)}{3} \right ) \right \}
  \langle \eta_c(p) | \overline{c} \gamma_{\mu}  \gamma_5 c
| 0 \rangle \langle K(q)| \overline{s} \gamma^{\mu}   b | B(p+q) \rangle   \, ,
\end{eqnarray}
where the meson momenta are explicitly specified and chosen as
$p^2 = m_{ \eta_c}^2$. The $\eta_c$ meson decay constant is
defined by the relation,
\begin{equation}
\langle \eta_c(p) | \overline{c}
(0) \gamma_{\mu}\gamma_5 c(0) |0\rangle = -if_{\eta_c} p_{\mu}.
\end{equation}
The decay constant $f_{\eta_c}$ can be estimated from the QCD sum
rules  approach with the current $J_\alpha=\bar{c}\gamma_\alpha
\gamma_5 c$ \cite{SVZ} or phenomenological potential models, in
fact, the values obtained from those approaches do not differ from
each other much; new estimation based on the nonperturbative
approach of coupled Schwinger-Dyson equation and Bethe-Salpeter
equation is in preparation. We use the value obtained from the
potential model in this article \cite{constant}.

The $B-K$ form factor can be parameterized as
\begin{equation}
\langle K(q)| \overline{s} \gamma_{\mu} b | B(p+q) \rangle =  (2 q + p)_{\mu}  F^+_{BK}(p^2) -
\frac{ m_B^2 - m_K^2 }{p^2} p_{\mu} (F^+_{BK}(p^2) - F^0_{BK}(p^2) ),
\end{equation}
the above form factors $F^+_{BK}(p^2),F^0_{BK}(p^2) $ can be estimated from the light-cone
sum rules approach \cite{BKR,KRWWY,PBall}, here we take the value
\begin{equation}
F^0_{BK}(m_{\eta_c}^2) = 0.42 \pm 0.06 \, .
\end{equation}
The concise expression for the factorizable matrix elements in the
decay $B\rightarrow \eta_c K$ can be written as
\begin{eqnarray}
 &&\langle \eta_c(p) K(q) |  H_w | B(p+q) \rangle=\frac{G_F}{\sqrt{2}}\left\{
 V_{c b} V_{c s}^* \left ( C_2(\mu) + \frac{C_1(\mu)}{3} \right ) \right.\nonumber\\
&& \ \ \ \left.-V_{t b} V_{t s}^*\left( C_3(\mu)-C_5(\mu) -C_7(\mu)+C_9(\mu)+ \frac{C_4(\mu)-C_6(\mu)-C_8(\mu)+C_{10}(\mu)}{3}\right)  \right \} if_{\eta_c}m^2_B F^0(m^2_{\eta_c}).
\end{eqnarray}
There are no factorizable contributions in the decay $B\rightarrow
\chi_{c0} K$  as the vector current $\bar{c}\gamma_\mu c$  is conserved and has a vanishing matrix
element with the $\chi_{c0}$ meson,
\begin{eqnarray}
\langle \chi_{c0}(p)|\bar{c}
c|0\rangle=f_{\chi_{c0}}m_{\chi_{c0}} \, ,\nonumber\\
 \langle \chi_{c0}(p) K(q) |  H_w | B(p+q) \rangle=0 \, ,
\end{eqnarray}
here $f_{\chi_{c0}}$ and $m_{\chi_{c0}}$ are the decay constant
and mass of the $\chi_{c0}$ meson respectively.

\section{ Light-cone sum rules for $\langle \eta_c K | {\cal \widetilde{O}} | B \rangle $ and $\langle \chi_{c0} K| {\cal \widetilde{O}} | B \rangle$}

In the following, we apply the approach developed in
Ref.\cite{Kh2001} for the $B \rightarrow \pi \pi$ channel to
estimate the contributions from the soft-gluon exchanges  in the
$B \rightarrow \eta_c K,\chi_{c0} K$ decays. Firstly, let us write
down the correlation functions,
\begin{eqnarray}
F^{\eta_c}_{\rho}(p,q,k) &=& i^2 \, \int d^4 x e^{-i(p-q)x} \int d^4 y e^{i(p-k)y} \langle 0 |
T \{ j_{\rho}^{\eta_c}(y) {\cal \widetilde{O}}(0) j_5^{B}(x) \} | K(q) \rangle \, , \\
F_{\chi_{c0}}(p,q,k) &=& i^2 \, \int d^4 x e^{-i(p-q)x} \int d^4 y e^{i(p-k)y} \langle 0 |
T \{ j^{\chi_{c0}}(y) {\cal \widetilde{O}}(0) j_5^{B}(x) \} | K(q) \rangle \, ,
\end{eqnarray}
where $j_{\rho}^{\eta_c} = \overline{c} \gamma_{\rho}\gamma_5 c$,
$j^{\chi_{c0}} = \overline{c}  c$ and  $j_5^{B} = m_b \overline{b}
i \gamma_5 u$ are currents interpolating the $\eta_c$ ,$\chi_{c0}$
and $B$ meson fields, respectively. We take the same notation
${\cal \widetilde{O}}$ for the operators ${\cal \widetilde{O}}_i$
as our final  result indicates that they have the same analytical
expressions due to their special Dirac structures for  the decays
$B \longrightarrow \eta_c K ,\chi_{c0}K $ .

The correlation functions can be calculated  by the operator
product expansion method near the light-cone $x^2\sim y^2\sim
(x-y)^2\sim 0$ in  QCD. They are functions of three
independent momenta chosen to be $q$, $p-k$ and $k$ by
convenience. Here we introduce the unphysical  momentum $k$ in
order to avoid that the $B$ meson has the same four-momentum
before ($p-q$) and after the decay ($P$).
In such a way, we can avoid a continuum of light contributions in the
dispersion relation in the $B$-channel.
The independent
kinematical invariants can be taken as $(p-q)^2$, $(p-k)^2$, $q^2$
, $k^2$,
 $P^2 = (p-k-q)^2$ and $p^2$. We set
$k^2 =0$ and take $q^2 = m_K^2 =0$, neglecting the small corrections of the order $O(m_K^2/m_B^2)$.
The momentum $p^2$  is  kept undefined for the moment  in order to   make the
derivation of the sum rules without restriction. Its value is going to be set later in this section, and chosen
$p^2 = m_{\eta_c}^2,m_{\chi_{c0}}^2$.
The values of $(p-k)^2$, $(p-q)^2$ and $P^2$ should be spacelike
and large in order to stay far away from the hadronic thresholds
in the $B$, $\eta_c$ and $\chi_{c0}$ channels. All together, we
have
\begin{eqnarray}
q^2 = k^2 = 0, \; p^2  (undefined), \; |(p-k)|^2 \gg \Lambda_{QCD}, |(p-q)|^2 \gg\Lambda_{QCD},  |P|^2 \gg \Lambda_{QCD} \, \, .
\,  \nonumber
\end{eqnarray}
 The decomposition of the
correlation function in Eq.(11) with the independent
momenta are straightforward and it can be
divided into the following Lorentz invariant
amplitudes\footnote{ $F_{\eta_c}$ and $F_{\chi_{c0}}$ are
functions of $q$, $p-k$ and $k$, their momenta dependence will be
written explicitly when necessary, for example, $F_{\eta_c}(p-k)$
and $F_{\chi_{c0}}(p-k)$, with emphasis on the dependence on the momentum $p-k$.  },
\begin{eqnarray}
F_\rho^{\eta_c}(p,q,k)&=&
(p-k)_\rho F_{\eta_c}(p,q,k)
+ q_\rho \widetilde{F}^{\eta_c}_1
+ k_\rho \widetilde{F}^{\eta_c}_2
+ \epsilon_{\rho \beta\lambda\xi}q^\beta p^\lambda k^\xi
\widetilde{F}^{\eta_c}_3\, .
\end{eqnarray}
According to the basic assumption of current-hadron duality in the
QCD sum rules  approach \cite{SVZ}, we insert  a complete series of
intermediate states satisfying the unitarity   principle with the same quantum numbers as the current
operators $j_{\rho}^{\eta_c}$, $j^{\chi_{c0}}$ and $j_5^{B} $
into the correlation functions in Eq.(11) and Eq.(12) to obtain
the hadronic representation. After isolating the pole terms of the
lowest pseudoscalar $\eta_c$ and $\chi_{c0}$  mesons in the
charmonium channels, we get the following result,
\begin{eqnarray}
F^{\eta_c}_{\rho}(p,q,k) &=&   \frac{\langle 0 |
j_{\rho}^{\eta_c}(0) |\eta_c (p-k)
\rangle}{m^2_{\eta_c}-(p-k)^2-i\varepsilon} \Pi_{\eta_c}( (p-q)^2,
P^2, p^2) + \int_{s_0^{\eta_c}}^{\infty} ds
\frac{\rho_{\rho}^{\eta_c}(s,(p-q)^2,P^2,p^2)}
{s - (p-k)^2}; \\
F_{\chi_{c0}}(p,q,k) &=&   \frac{\langle 0 | j^{\chi_{c0}}(0)
|\chi_{c0} (p-k) \rangle}{m^2_{\chi_{c0}}-(p-k)^2-i\varepsilon}
\Pi_{\chi_{c0}}( (p-q)^2, P^2, p^2) +
\int_{s_0^{\chi_{c0}}}^{\infty} ds
\frac{\rho^{\chi_{c0}}(s,(p-q)^2,P^2,p^2)} {s - (p-k)^2},
\end{eqnarray}
where
\begin{eqnarray}
\Pi_{\eta_c(\chi_{c0})}( (p-q)^2, P^2, p^2)&=& i \int d^4 x e^{-i(p-q)x}
\langle \eta_c (\chi_{c0})(p-k) K(-q) |
T \{ {\cal O}(0) j_5^{B}(x) \}|0 \rangle \,  .
\end{eqnarray}
In above equations, $\rho_{\rho}^{\eta_c}$ and $s_0^{\eta_c}$  are
the spectral density and threshold parameter  of the lowest
excited resonances and  continuum states in the $\eta_c$ channel,
respectively; while $\rho^{\chi_{c0}}$  and $s_0^{\chi_{c0}}$ are
the corresponding ones in the $\chi_{c0}$ channel. In the limit of
large spacelike momentum  $(p-k)^2 \ll
m_{\eta_c}^2,m^2_{\chi_{c0}}$, the correlation functions in
Eq.(11) and Eq.(12) can be calculated in  QCD at the
level of quark-gluon degrees of freedom and rewritten in the
following forms by applying  dispersion relation,
\begin{eqnarray}
F^{\eta_c}_{\rho}(F_{\chi_{c0}})& =& \frac{1}{\pi}
\int_{4 m_c^2}^{\infty} ds \frac{{\rm Im}_s F^{\eta_c}_{\rho}(F_{\chi_{c0}})
(s,(p-q)^2,P^2,p^2)}{s - (p-k)^2} \, .
 \end{eqnarray}
We can approximate the hadronic spectral densities
$\rho_{\rho}^{\eta_c}$ and $\rho^{\chi_{c0}}$ above the thresholds
of the lowest
excited resonances and continuum  states by the corresponding ones from  QCD
calculations with  the assumption of quark-hadron duality,
\begin{eqnarray}
\rho_{\rho}^{\eta_c}(\rho^{\chi_{c0}})(s,(p-q)^2,P^2,p^2)\Theta(s-s_0^{\eta_c}(s_0^{\chi_{c0}})) =
\frac{1}{\pi} {\rm Im}_s F^{\eta_c}_{\rho}(F_{\chi_{c0}})(s,(p-q)^2,P^2,p^2)\Theta(s - s_0^{\eta_c}(s_0^{\chi_{c0}})) \, .
\end{eqnarray}
Now we explore the analytical properties of the amplitudes
$\Pi_{\eta_c}((p-q)^2, P^2, p^2)$, $\Pi_{\chi_{c0}}((p-q)^2, P^2,
p^2)$ in the $B$-channel and insert  a complete series of
 hadronic states with the same quantum numbers as the $B$ meson into the correlation functions in
 Eq.(16). After isolating  the lowest pole terms of the
B meson contributions, we obtain the following results,
\begin{eqnarray}
  \Pi_{\eta_c(\chi_{c0})}((p-q)^2, P^2, p^2  ) &=&  \frac{m_{B}^2 f_{B}}{m_{B}^2 - (p-q)^2-i\varepsilon}
\langle \eta_c(\chi_{c0})(p-k) K(-q) | {\cal O}(0) | B(p+q) \rangle
 + \int_{s_0^{B}}^{\infty} ds' \frac{\rho^{B}_{\eta_c(\chi_{c0})}(s',P^2,p^2)}{s' - (p-q)^2} \, .
 \end{eqnarray}
 The hadronic spectral densities
$\rho^{B}_{\eta_c}$ and $\rho^{B}_{\chi_{c0}}$ above the threshold
of the continuum $s_0^{B}$ can be approximated by the corresponding ones at the
level of quark-gluon degrees of freedom.

Finally, we obtain the correlation functions in hadronic
representation,
\begin{eqnarray}
F_{\rho}^{\eta_c}(p,q,k) &=& \frac{\langle 0 |
j_{\rho}^{\eta_c}(0) |\eta_c (p-k)
\rangle}{m^2_{\eta_c}-(p-k)^2-i\varepsilon} \langle \eta_c
(p-k)|{\cal\widetilde{ O}}(0)|B(p-q)\rangle | K(q) \rangle
 \frac{\langle B(p-q|  j_5^{B}(0)  | 0 \rangle}{m^2_{B}-(p-q)^2-i\varepsilon}  +\cdots,  \nonumber\\
&=& \frac{i f_{\eta_c}(p-k)_{\rho}}{m^2_{\eta_c}-(p-k)^2-i\varepsilon}\frac{f_B m^2_B}{m^2_{B}-(p-q)^2-i\varepsilon}
\langle \eta_c (p-k)K(-q)|{\cal \widetilde{O}}(0)|B(p-q)\rangle+\cdots , \nonumber \\
&=&(p-k)_\rho F_{\eta_c}(p,q,k)+\cdots; \\
F_{\chi_{c0}}(p,q,k) &=& \frac{\langle 0 | j^{\chi_{c0}}(0)
|\chi_{c0} (p-k) \rangle}{m^2_{\chi_{c0}}-(p-k)^2-i\varepsilon}
\langle \chi_{c0} (p-k)|{\cal\widetilde{ O}}(0)|B(p-q)\rangle |
K(q) \rangle
 \frac{\langle B(p-q|  j_5^{B}(0)  | 0 \rangle}{m^2_{B}-(p-q)^2-i\varepsilon}
 +\cdots,
  \nonumber\\
&=& \frac{f_{\chi_{c0}}m_{\chi_{c0}}}{m^2_{\chi_{c0}}-(p-k)^2-i\varepsilon}\frac{f_B m^2_B}{m^2_{B}-(p-q)^2-i\varepsilon}
\langle \eta_c (p-k)K(-q)|{\cal \widetilde{O}}(0)|B(p-q)\rangle+\cdots .
\end{eqnarray}
Here we do not show the contributions from the higher resonances
and continuum states above the thresholds explicitly , they can be written in
terms of dispersion integrals and the spectral densities
 can be approximated by the quark-hadron duality ansatz. In
Eq.(20), we  select the relevant terms with tensor structure
$(p-k)_{\rho}$, which corresponding to the contributions from the
pseudoscalar mesons, for example, the $\eta_c$ and B mesons.

In order to suppress the contributions from the excited and
continuum states in the charmonium channels, we can perform n-th
derivative with respect to the momentum $(p-k)^2$ in Eqs.(14-15) to obtain
  n-th moment sum rules for the correlation functions
$\Pi_{\eta_c}((p-q)^2, P^2, p^2)$ and $\Pi_{\chi_{c0}}((p-q)^2, P^2,
p^2)$ in hadronic representation,
\begin{eqnarray}
i(p-k)_\rho \Pi_{\eta_c}((p-q)^2, P^2, p^2 ) &=& \frac{1}{\pi \,
f_{_{\eta_c}}} \int_{4 m_c^2}^{s_0^{_{\eta_c}}} ds
\frac{(m_{_{\eta_c}}^2 + Q_0^2)^{n+1}}{(s + Q_0^2)^{n+1}}
{\rm Im}_s { F}^{\eta_c}_{\rho}(s,(p-q)^2,P^2, p^2) \, , \nonumber\\
&=&\frac{1}{\pi^2 \,  f_{_{\eta_c}}} \int_{4
m_c^2}^{s_0^{_{\eta_c}}} ds \frac{(m_{_{\eta_c}}^2 +
Q_0^2)^{n+1}}{(s + Q_0^2)^{n+1}}
\int_{m_b^2}^{\infty}\frac{ds'}{s'-(p-q)^2}
{\rm Im}_s {\rm Im}_{s'}{ F}^{\eta_c}_{\rho}(s,s',P^2, p^2) \, ; \\
\Pi_{\chi_{c0}}((p-q)^2, P^2, p^2 ) &=& \frac{1}{\pi \,
m_{_{\chi_{c0}}} f_{_{\chi_{c0}}}} \int_{4
m_c^2}^{s_0^{_{\chi_{c0}}}} ds \frac{(m_{_{\chi_{c0}}}^2 +
Q_0^2)^{n+1}}{(s + Q_0^2)^{n+1}}
{\rm Im}_s { F}_{\chi_{c0}}(s,(p-q)^2,P^2, p^2) \, , \nonumber\\
&=& \frac{1}{\pi^2 \, m_{_{\chi_{c0}}} f_{_{\chi_{c0}}}} \int_{4
m_c^2}^{s_0^{_{\chi_{c0}}}} ds \frac{(m_{_{\chi_{c0}}}^2 +
Q_0^2)^{n+1}}{ (s +
Q_0^2)^{n+1}}\int_{m_b^2}^{\infty}\frac{ds'}{s'-(p-q)^2} {\rm
Im}_s{\rm Im}_{s'} { F}_{\chi_{c0}}(s,s',P^2, p^2) \, ,
 \end{eqnarray}
where the imaginary parts with respect to $s$ and $s'$ are given
by
\begin{eqnarray}
{\rm Im}_s {\rm Im}_{s'}{ F}^{\eta_c}_{\rho}({
F}_{\chi_{c0}})(s,s',P^2, p^2)&=&i(p-k)_\rho
f_{\eta_c}(f_{\chi_{c0}}m_{\chi_{c0}}) \pi^2 f_B m^2_B\langle
\eta_c(\chi_{c0}) (p-k)K(-q)| {\cal
\widetilde{O}}(0)|B(p-q)\rangle \nonumber \\
&&\delta(s'-m^2_B)\delta(s-m^2_{\eta_c (\chi_{c0})})+\cdots \, .
\end{eqnarray}
Here $Q_0$ is the  parameter for QCD sum rules in the charmonium
channels, the spectral densities above the thresholds can be approximated by the corresponding ones from
QCD calculation and not shown explicitly.

The Borel transformations with respect to $(p-q)^2$ in the B
channels in Eqs.(22-23) are straightforward.  Comparing  with
Eqs.(14-16), we can obtain the following results\footnote{Here we
prefer the notations $F_{\eta_c}(p-k),F_{{\chi_{c0}}}(p-k)$ to
$F_{\eta_c}(p,q,k),F_{{\chi_{c0}}}(p,q,k)$, with emphasis on the
dependence on $p-k$.   },
\begin{eqnarray}
 B_{trans}F_{\eta_c(\chi_{c0})}(p-k) &=& i f_{\eta_c}(f_{ \chi_{c0}}m_{\chi_{c0}})f_B m^2_B\langle
\eta_c(\chi_{c0}) (p-k)K(-q)| {\cal
\widetilde{O}}(0)|B(p-q)\rangle
 \frac{1}{(m^2_{\eta_c(\chi_{c0})} + Q_0^2)^{n+1}} \frac{e^{-\frac{m_B^2}{M^2}}}{M^2}+\cdots \, ,
 \end{eqnarray}
here  $B_{trans}$ denotes both the n-th derivative and Borel
transformation,   $M^2$ is the Borel parameter in the B channel.
The contributions from the excited and continuum states are not
shown explicitly for simplicity.

Then we can analytically continue $P^2$ from the space-like region
$P^2\ll 0$ to the time-like region  $P^2 \ge 0$,  and choose $P^2 =
m_B^2$. Now we carry out the operator product expansion near
the light-cone to obtain the    representation at the level of
quark-gluon degrees of freedom for the amplitudes $F_{\eta_c}$ and
$F_{{\chi_{c0}}}$.
 Firstly, let us write down
the propagator of a massive quark in the external gluon field
in the Fock-Schwinger gauge\cite{BBKR,BK},
\begin{eqnarray}
\langle 0 | T \{q_i(x_1)\, \bar{q}_j(x_2)\}| 0 \rangle
& =&i \int\frac{d^4k}{(2\pi)^4}e^{-ik(x_1-x_2)}\Bigg\{
\frac{\not\!k +m}{k^2-m^2} \delta_{ij}
-\int\limits_0^1 dv\,  g_s \, G^{\mu\nu}_a(vx_1+(1-v)x_2)
\left (\frac{\lambda^a}{2} \right )_{ij}
\nonumber
\\
& &  \Big[ \frac12 \frac {\not\!k +m}{(k^2-m^2)^2}\sigma_{\mu\nu} -
\frac1{k^2-m^2}v(x_1-x_2)_\mu \gamma_\nu \Big]\Bigg\}\, ,
\end{eqnarray}
here $G^{\mu \nu }_a$ is the gluonic field strength, $g_s$ denotes the strong
coupling constant.

Substituting the above b and c quark propagators  into the
correlation functions in Eqs.(11-12), we can obtain the hadronic
spectral densities  at the level of quark-gluon degrees of
freedom. The following three particle kaon distribution amplitudes
are useful in our calculation,
\\
- {\it  twist-3 distribution amplitude}
\begin{eqnarray}
 \langle 0 |\overline{s}(0) \sigma_{\mu \nu} \gamma_5 G_{\alpha \beta}(v y) u(x) | K^{+}(q) \rangle &=&
i f_{3 K} \left [ (q_{\alpha}q_{\mu}g_{\beta \nu} - q_{\beta}q_{\mu}g_{\alpha \nu}) \right .  \nonumber \\
& &     \left . - (q_{\alpha}q_{\nu}g_{\beta \mu} - q_{\beta}q_{\nu}g_{\alpha \mu}) \right ]
\int {\cal D}\alpha_i \phi_{3 K}(\alpha_i, \mu)
e^{-i q(x \alpha_1 + y v \alpha_3)} \, ;
\end{eqnarray}
- {\it twist-4 distribution amplitudes}
\begin{eqnarray}
 \langle 0 |\overline{s}(0) i\gamma_{\mu} \tilde{G}_{\alpha \beta}(v y) u(x) | K^{+}(q) \rangle &=&
q_{\mu} \frac{q_{\alpha} x_{\beta} - q_{\beta}x_{\alpha}}{q x} f_{K} \int {\cal D}\alpha_i \tilde{\phi}_{\parallel}(\alpha_i,\mu)
e^{-i q(x \alpha_1 + y v \alpha_3)} \nonumber \\
& &  +
(g_{\mu \alpha}^{\perp}q_{\beta} - g_{\mu \beta}^{\perp}q_{\alpha}) \int {\cal D}\alpha_i \tilde{\phi}_{\perp}(\alpha_i,\mu)
e^{-i q(x \alpha_1 + y v \alpha_3)} \, ;
\end{eqnarray}
\begin{eqnarray}
 \langle 0 |\overline{s}(0) \gamma_{\mu} \gamma_5 {G}_{\alpha \beta}(v y) u(x) |K^{+}(q) \rangle &=&
q_{\mu} \frac{q_{\alpha} x_{\beta} - q_{\beta}x_{\alpha}}{q x} f_{K} \int {\cal D}\alpha_i {\phi}_{\parallel}(\alpha_i,\mu)
e^{-i q(x \alpha_1 + y v \alpha_3)} \nonumber \\
& & +
(g_{\mu \alpha}^{\perp}q_{\beta} - g_{\mu \beta}^{\perp}q_{\alpha}) \int {\cal D}\alpha_i {\phi}_{\perp}(\alpha_i,\mu)
e^{-i q(x \alpha_1 + y v \alpha_3)} \, ,
\end{eqnarray}
where
\begin{eqnarray}
&&\tilde{G}_{\alpha \beta} = \frac{1}{2} \epsilon_{\alpha \beta \rho \sigma} G^{\rho \sigma}, \ \ \
G^{\rho \sigma} = g_s \, \frac{\lambda^{a} G^{\rho \sigma}_a}{2} \, ; \nonumber\\
&& {\cal D} \alpha_i =
d\alpha_1 d\alpha_2 d\alpha_3 \delta( 1 - \alpha_1 - \alpha_2 - \alpha_3), \ \ \  g_{\alpha \beta}^{\perp} =
g_{\alpha \beta} - \frac{x_{\alpha} q_{\beta} + x_{\beta} q_{\alpha}}{qx}.
\end{eqnarray}
The twist-3 and twist-4 light-cone distribution amplitudes can be parameterized as
\begin{eqnarray}
\phi_{3K}(\alpha_i,\mu) &=& 360 \alpha_1 \alpha_2 \alpha_3^2 \left (1 + a(\mu) \frac{1}{2} ( 7 \alpha_3 - 3) + b(\mu)
(2 - 4 \alpha_1 \alpha_2 - 8 \alpha_3 (1 - \alpha_3)) \right . \nonumber \\
& &  + \left . c(\mu) ( 3 \alpha_1 \alpha_2 - 2 \alpha_3 + 3 \alpha_3^2) \right ) \, ,
 \\
\phi_{\perp}(\alpha_i,\mu) &=& 30 \delta^2(\mu)(\alpha_1-\alpha_2)\alpha_3^2\left [ \frac{1}{3} +
2 \epsilon (\mu) (1 - 2 \alpha_3) \right ]  \, ,
 \\
\phi_{\parallel}(\alpha_i,\mu) &=& 120 \delta^2(\mu) \epsilon (\mu)  (\alpha_1-\alpha_2) \alpha_1 \alpha_2 \alpha_3  \, ,
\\
\tilde{\phi}_{\perp}(\alpha_i,\mu) &=&
30 \delta^2(\mu) \alpha_3^2 ( 1 - \alpha_3) \left [ \frac{1}{3} + 2 \epsilon (\mu)  (1 - 2 \alpha_3) \right ] \, ,
 \\
\tilde{\phi}_{\parallel}(\alpha_i,\mu) &=& -120 \delta^2(\mu) \alpha_1 \alpha_2 \alpha_3 \left [ \frac{1}{3} +
\epsilon (\mu) (1 - 3 \alpha_3) \right ] \, .
 \
\end{eqnarray}
Those parameters in the light-cone distribution amplitudes can be
estimated with the QCD sum rules approach \cite{Kho1,CZ,BF}. In
practical manipulation, we can neglect the $a(\mu),b(\mu), c(\mu)$
and $\epsilon(\mu)$ dependence and the asymptotic forms  will be taken.

After carrying out the operator product expansion near the light-cone,
we obtain the following expressions for the $F_{\eta_c}$ and
$F_{\chi_{c0}}$ ,
\begin{eqnarray}
F_{\eta_c}(p-k) &=&  \frac{m_b f_{3 K}}{4 \pi^2} \int_0^1 dv \int {\cal D}\alpha_i \frac{\phi_{3K}(\alpha_i, \mu)}
{m_b^2 - (p - q (1-\alpha_1))^2} \,
\int_0^1 dx \frac{ 2 x^2 (1-x)}{ m_c^2 - (p-k-v \alpha_3 q)^2 x (1-x)
} \nonumber \\
& &  q \cdot (p-k) \Big [ (2-v) q \cdot k + 2 (1-v) q \cdot (p-k) \Big ] \nonumber \\
&&-\frac{m^2_b f_{ K}}{4 \pi^2} \int_0^1 dv \int {\cal D}\alpha_i \frac{\tilde{\phi}_{\parallel}(\alpha_i,\mu) +\tilde{\phi}_{\perp}(\alpha_i,\mu)}
{m_b^2 - (p - q (1-\alpha_1))^2}\int_0^1 dx \frac{ 2 x^2 (1-x)q\cdot(p-k)}{ m_c^2 - (p-k-v \alpha_3 q)^2 x (1-x)
} \nonumber \\
&&-\frac{m^2_b f_{ K}}{4 \pi^2} \int_0^1 dv \int {\cal D}\alpha_i \frac{\tilde{\phi}_{\perp}(\alpha_i,\mu)}
{m_b^2 - (p - q (1-\alpha_1))^2}\int_0^1 dx \frac{ 2 x^2 (1-x)(2v-3)q\cdot(p-k)}{ m_c^2 - (p-k-v \alpha_3 q)^2 x (1-x)
} \, ; \\
F_{\chi_{c0}}(p-k) &=&  \frac{m_b m_cf_{3 K}}{4 \pi^2} \int_0^1 dv \int {\cal D}\alpha_i \frac{\phi_{3K}(\alpha_i, \mu)}
{m_b^2 - (p - q (1-\alpha_1))^2} \,
\int_0^1 dx \frac{ (2 x^2 -2x(1-x)(1-2v))q \cdot (p-k)  q\cdot p}{ m_c^2 - (p-k-v \alpha_3 q)^2 x (1-x)
} \nonumber \\
&&-\frac{m^2_b m_cf_{ K}}{4 \pi^2} \int_0^1 dv \int {\cal D}\alpha_i \frac{(\phi_{\parallel}(\alpha_i,\mu) -2\phi_{\perp}(\alpha_i,\mu))}
{m_b^2 - (p - q (1-\alpha_1))^2}\int_0^1 dx \frac{  (x^2- x(1-x)(1-2v))q\cdot(p-k)}{ m_c^2 - (p-k-v \alpha_3 q)^2 x (1-x)
} \, .
\end{eqnarray}

Here we will take a short digression to discuss the technical
details. In calculation, we will encounter x-integral of the form
\begin{eqnarray}
\int d^4 x \frac{x_\rho}{q \cdot x} f(p,q,k,x) \nonumber
\end{eqnarray}
in the coordinates  representation, which can be formally written as
\begin{eqnarray}
\int d^4 x \frac{x_\rho}{q \cdot x} f(p,q,k,x)=A(p,q,k)(p-k)_\rho
+ B(p,q,k)q_\rho  + C(p,q,k)k_\rho  + \epsilon_{\rho
\beta\lambda\xi}q^\beta p^\lambda k^\xi D(p,q,k)\, .
\end{eqnarray}
Multiplying both sides of above equation by $q_\rho$ and taking
the Chiral limit $q^2=m_K^2=0$, the expression of the relevant
quantity $A(p,q,k)$ can be obtained.

 It is easy to perform the Feynman parameter  $x$ integral in
 Eqs.(36-37),  we prefer this form in order to facilitate the Borel
transformation and n-th derivative. For the case of  massless
quark loops, one can integrate out the variable $x$ directly.

In the following, we write down the dispersion  relations
for the correlation functions at the level of quark-gluon degrees
of freedom,
\begin{eqnarray}
F_{\eta_c}(p-k) &=&  \frac{m_b }{4 \pi^2} \int_0^1 dv \int {\cal D}\alpha_i
\int_{m^2_b}^{s_{B}}d s_1 \int_{4m^2_c}^{s_{\eta_c}}d s_2 \int_0^1 dx\frac{ x^2 (1-x)}
{s_1- (p - q)^2} \,
 \frac{s_2-P^2 }{ s_2 - (p-k)^2
} \nonumber \\
& & \left\{f_{3 K}\phi_{3K}(\alpha_i, \mu)\Big [ \frac{2-v}{2}(P^2+m^2_{\eta_c}-s_1-s_2) + (1-v) (s_2-P^2) \Big ]\right.\nonumber\\
&&\left.-m_b f_{ K}  (\tilde{\phi}_{\parallel}(\alpha_i,\mu) +\tilde{\phi}_{\perp}(\alpha_i,\mu))
-m_b f_{ K}\tilde{\phi}_{\perp}(\alpha_i,\mu)(2v-3) \right \}  \nonumber\\
&&\delta(m^2_b-m^2_{\eta_c}\alpha_1-s_1 (1-\alpha_1)) \delta(m_c^2-x(1-x)v\alpha_3 P^2-x(1-x)(1-v\alpha_3)s_2)+\cdots \, ; \\
F_{\chi_{c0}}(p-k) &=&  \frac{m_bm_c }{8 \pi^2} \int_0^1 dv \int {\cal D}\alpha_i
\int_{m^2_b}^{s_{B}}d s_1 \int_{4m^2_c}^{s_{\chi_{c0}}}d s_2 \int_0^1 dx \frac{x^2- x(1-x)(1-2v)}{s_1- (p - q)^2} \,
 \frac{s_2-P^2 }{ s_2 - (p-k)^2
}  \nonumber \\
& &  \left\{ f_{3 K}\phi_{3K}(\alpha_i, \mu)(m^2_{\chi_{c0}}-s_1)-m_b f_{ K}
(\phi_{\parallel}(\alpha_i,\mu) -2\phi_{\perp}(\alpha_i,\mu))\right\} \nonumber\\
&&\delta(m^2_b-m^2_{\chi_{c0}}\alpha_1-s_1 (1-\alpha_1)) \delta(m_c^2-x(1-x)v\alpha_3 P^2-x(1-x)(1-v\alpha_3)s_2)+\cdots.
\end{eqnarray}
Again, the higher resonances and continuum states contributions are not
shown explicitly for simplicity, as they are Borel transformation
or n-th derivative suppressed. Here we can  introduce some
notations to simplify the cumbersome expressions in Eq.(39) and
Eq.(40) respectively,
\begin{eqnarray}
x_i=\frac{1}{2} \left(1-\sqrt{1-\frac{4m_c^2}{s}} \right) \, , & \ \ \ & x_f=\frac{1}{2} \left(1+\sqrt{1-\frac{4m_c^2}{s}} \right)\, , \nonumber \\
\alpha_c=\frac{x(1-x)s-m_c^2}{x(1-x)(s-P^2)} \, , & \ \ \ & \alpha_0=\frac{m_b^2-m^2_{\eta_c}(m^2_{\chi_{c0}})}{s_B-m^2_{\eta_c}(m^2_{\chi_{c0}})}  \,.
\end{eqnarray}

In performing the $\delta$ functions integrals in Eq.(39) and
Eq.(40), we note that in the  space-like region $P^2 \ll  0$,  the
condition $\alpha_0 > \alpha_c$ can be warranted.
   Performing   Borel transformation in the B channel and
  n-th derivative
in the $\eta_c$, $\chi_{c0}$ channels, then matching with Eq.(25),
finally we obtain the sum rules for the nonfactorizable soft
matrix elements,
\begin{eqnarray}
\langle \eta_c (p-k)K(-q)|{\cal \widetilde{O}}(0)|B(p-q)\rangle
&=& \frac{im_b}{4\pi^2f_B f_{\eta_c} m^2_B}  \int_{4 m^2_c}^{s_{\eta_c}}d s
\int_{x_i}^{x_f}xdx
\int_{\alpha_0}^1 d\alpha
\int_{\alpha_c}^{\alpha} d \beta \left[ f_{3K}\phi_{3K}(1-\alpha,\alpha-\beta,\beta) \right. \nonumber\\
&& \left.\left\{ (s-P^2)(1-\frac{\alpha_c}{\beta}) -(1-\frac{\alpha_c}{2\beta})\frac{m_b^2-m^2_{\eta_c}(1-\alpha)+\alpha(s-P^2-m^2_{\eta_c}
)}{\alpha} \right\} \right.  \nonumber \\
&& - m_b f_K(\tilde{\phi}_{\parallel}(1-\alpha,\alpha-\beta,\beta) +\tilde{\phi}_{\perp}(1-\alpha,\alpha-\beta,\beta)) \nonumber\\
&& \left.-m_b f_K \tilde{\phi}_{\perp}(1-\alpha,\alpha-\beta,\beta)(\frac{2\alpha_c}{\beta}-3) \right] \nonumber\\
&&Exp(\frac{m_B^2 \alpha+m^2_{\eta_c}(1-\alpha)-m^2_b}{M^2 \alpha}) \left(\frac{m^2_{\eta_c}+Q_0^2}{s+Q_0^2}\right)^{n+1}\frac{1}{\alpha\beta }\, \, ;\\
\langle \chi_{c0} (p-k)K(-q)|{\cal \widetilde{O}}(0)|B(p-q)\rangle
&=& \frac{m_bm_c}{8\pi^2f_B f_{\chi_{c0}} m^2_B m_{\chi_{c0}}} \int_{4
m^2_c}^{s_{\chi_{c0}}}d s \int_{x_i}^{x_f}dx \int_{\alpha_0}^1 d\alpha
\int_{\alpha_c}^{\alpha} d \beta \left[  f_{3K} \phi_{3K}
 (1-\alpha,\alpha-\beta,\beta) \right. \nonumber\\
&&  \frac{m^2_{\chi_{c0}}-m^2_b}{\alpha}
-\left. m_b f_{K} (\phi_{\parallel}(1-\alpha,\alpha-\beta,\beta) -2\phi_{\perp}(1-\alpha,\alpha-\beta,\beta))\right]  \nonumber\\
&& \left\{ x -(1-x) \left(1-2 \frac{\alpha_c}{\beta }  \right) \right\}   \nonumber \\
&& Exp(\frac{m_B^2 \alpha+m^2_{\chi_{c0}}(1-\alpha)-m^2_b}{M^2
\alpha})
\left(\frac{m^2_{\chi_{c0}}+Q_0^2}{s+Q_0^2}\right)^{n+1}\frac{1}{\alpha\beta
(1-x) }.
\end{eqnarray}
In  above expressions, $P^2$ is chosen to be large space-like
squared momentum ($|P^2|\sim m_b^2$) in order to stay far away
from the hadronic thresholds in the channels of the B and
charmonia currents,
 the values of $\alpha_c$ are small positive quantities but not always small enough to be safely
  neglected, we can perform the following approximation for the $\beta$
integral,
\begin{eqnarray}
\int_{\alpha_c}^{\alpha} d\beta
G(s,x,\alpha,\beta)=\left\{\int^{\alpha}_{0}
-\int^{\alpha_c}_{0}\right\} d\beta G(s,x,\alpha,\beta),
\end{eqnarray}
here $G$ is an abbreviation for the integral functions and can be
written as
\begin{eqnarray}
G(s,x,\alpha,\beta)=A(s,x,\alpha,\beta)\phi_{3K}(1-\alpha,\alpha-\beta,\beta)
+B(s,x,\alpha,\beta)\tilde{\phi}_{\parallel}(1-\alpha,\alpha-\beta,\beta)
+C(s,x,\alpha,\beta)\tilde{\phi}_{\perp}(1-\alpha,\alpha-\beta,\beta)
\nonumber
\end{eqnarray}
or
\begin{eqnarray}
G(s,x,\alpha,\beta)=D(s,x,\alpha,\beta)\phi_{3K}(1-\alpha,\alpha-\beta,\beta)+
E(s,x,\alpha,\beta)\phi_{\parallel}(1-\alpha,\alpha-\beta,\beta)
+F(s,x,\alpha,\beta)\phi_{\perp}(1-\alpha,\alpha-\beta,\beta),
\nonumber
\end{eqnarray}
$A,B,C,D,E,F$ are formal notations. We can expand the light-cone
distribution amplitudes $\phi_{3K}$, $ \tilde{\phi}_{\parallel}$,
$\tilde{\phi}_{\perp}$, $\phi_{\parallel}$ and $\phi_{\perp}$ in
terms of Taylor series of $\beta$ \footnote{For very small
$\alpha_c$, we can approximate the integral
$\int_0^{\alpha_c}d\beta G(s,x,\alpha,\beta)$ by $
\alpha_cG(s,x,\alpha,\beta)|_{\beta\rightarrow 0 } $, then
analytically continue $P^2$ into the timelike region,
$P^2=m^2_B$.}, for example,
\begin{eqnarray}
\phi_{3K}(1-\alpha,\alpha-\beta,\beta)&=&\phi_{3K}(1-\alpha,\alpha-\beta,\beta)|_{\beta=0}+\frac{\partial}{\partial
\beta }\phi_{3K}(1-\alpha,\alpha-\beta,\beta)|_{\beta=0}\beta\nonumber\\
 &&+\frac{1}{2}\frac{\partial^2}{\partial \beta^2
}\phi_{3K}(1-\alpha,\alpha-\beta,\beta)|_{\beta=0}\beta^2+\cdots
\, ,
\end{eqnarray}
 and analytically continue $P^2$ into the
timelike region, $P^2=m^2_B$, then complete the integral $\int^{\alpha_c}_{0} d\beta
G(s,x,\alpha,\beta)$.  The explicit expressions for the
physical matrix elements $\langle \eta_c (p)K(-q)|{\cal
{\widetilde{O}}}(0)|B(p-q)\rangle$
 and $\langle \chi_{c0} (p)K(-q)|{\cal {\widetilde{O}}}(0)|B(p-q)\rangle$ are
 lengthy due to the re-summation of all the Taylor series of $\beta$,  here we show only the leading terms explicitly,
\begin{eqnarray}
\langle \eta_c (p)K(-q)|{\cal \widetilde{O}}(0)|B(p-q)\rangle &=&
\frac{m_bi}{4\pi^2f_B f_{\eta_c} m^2_B} \int_{4 m^2_c}^{s_{\eta_c}}d s
\int_{x_i}^{x_f}xdx
\int_{\alpha_0}^1 d\alpha  \nonumber\\
&& \left[  f_{3K}  \left\{ \int_{0}^{\alpha} d \beta \phi_{3K}(1-\alpha,\alpha-\beta,\beta)-\int_{0}^{\alpha_c} d \beta \phi_{3K}(1-\alpha,\alpha-\beta,\beta)|_{\beta=0} \right\}  \right.\nonumber\\
&& \left.\left\{ (s-m_B^2)(1-\frac{\alpha_c}{\beta})
-(1-\frac{\alpha_c}{2\beta})\frac{m_b^2-m^2_{\eta_c}(1-\alpha)+\alpha(s-m_B^2-m^2_{\eta_c}
)}{\alpha} \right\} \right.  \nonumber \\
&& \left.-m_b f_K
\left\{\int_{0}^{\alpha} d \beta  (\tilde{\phi}_{\parallel}(1-\alpha,\alpha-\beta,\beta) +\tilde{\phi}_{\perp}(1-\alpha,\alpha-\beta,\beta)) \right.\right.\nonumber\\
&&\left.\left.-\int_{0}^{\alpha_c} d \beta  (\tilde{\phi}_{\parallel}(1-\alpha,\alpha-\beta,\beta)|_{\beta=0} +\tilde{\phi}_{\perp}(1-\alpha,\alpha-\beta,\beta)|_{\beta=0}) \right\} \right.\nonumber\\
&&\left. -m_b f_K
\left\{\int_{0}^{\alpha} d \beta  \tilde{\phi}_{\perp}(1-\alpha,\alpha-\beta,\beta)-\int_{0}^{\alpha_c} d \beta  \tilde{\phi}_{\perp}(1-\alpha,\alpha-\beta,\beta)|_{\beta=0}\right\} \right.\nonumber\\
&&\left. (\frac{2\alpha_c}{\beta}-3) \right] Exp(\frac{m_B^2 \alpha+m^2_{\eta_c}(1-\alpha)-m^2_b}{M^2 \alpha}) \left(\frac{m^2_{\eta_c}+Q_0^2}{s+Q_0^2}\right)^{n+1}\frac{1}{\alpha\beta }+\cdots \, ,  \\
\langle \chi_{c0} (p)K(-q)|{\cal \widetilde{O}}(0)|B(p-q)\rangle
&=& \frac{m_b m_c}{8\pi^2 f_B f_{\chi_{c0}} m^2_B m_{\chi_{c0}}} \int_{4
m^2_c}^{s_{\chi_{c0}}}d s \int_{x_i}^{x_f}dx \int_{\alpha_0}^1 d\alpha\left[ f_{3K} \frac{m^2_{\chi_{c0}}-m^2_b}{\alpha}\right. \nonumber\\
&&\left\{\int_{0}^{\alpha} d \beta \phi_{3K}
 (1-\alpha,\alpha-\beta,\beta)-\int_{0}^{\alpha_c} d \beta \phi_{3K}
 (1-\alpha,\alpha-\beta,\beta)|_{\beta=0} \right\} \nonumber\\
&&-m_b f_{K}
\left\{\int_{0}^{\alpha} d \beta (\phi_{\parallel}(1-\alpha,\alpha-\beta,\beta)
-2\phi_{\perp}(1-\alpha,\alpha-\beta,\beta)) \right. \nonumber \\
&&\left.\left.-\int_{0}^{\alpha_c} d \beta (\phi_{\parallel}(1-\alpha,\alpha-\beta,\beta)|_{\beta=0}
-2\phi_{\perp}(1-\alpha,\alpha-\beta,\beta)|_{\beta=0})\right\} \right]\nonumber \\
&& \left\{ x -(1-x) \left(1-2 \frac{\alpha_c}{\beta }  \right) \right\}  \nonumber\\
&&Exp(\frac{m_B^2 \alpha+m^2_{\chi_{c0}}(1-\alpha)-m^2_b}{M^2
\alpha})
\left(\frac{m^2_{\chi_{c0}}+Q_0^2}{s+Q_0^2}\right)^{n+1}\frac{1}{\alpha\beta
(1-x) }+\cdots.
\end{eqnarray}
 In performing the $\beta$ integral $\int_0^{\alpha_c}$, we need only the
values of the light-cone distribution amplitudes $\phi_{3K}$, $
\tilde{\phi}_{\parallel}$, $\tilde{\phi}_{\perp}$,
$\phi_{\parallel}$, $\phi_{\perp}$ and their derivations at zero
momentum fraction i.e. $\beta=0$, there are no problems with
negative partons (quarks and gluons) momentum fractions. The analytical continuation of
 $P^2$ to its positive value ends up with an unavoidable
theoretical uncertainty, if only a few terms of the Taylor series
are taken, smaller $|\alpha_c|$ (In the Chiral limit $m_c=0$,
$|s/(s-m_B^2)|$) with greater precision, however, with the
re-summation to all orders of $\beta$ in Eq.(45), the
 assumption of quark-hadron duality is still applicable  in  the case of heavy
meson final states. This procedure ensures the  disappearance of
the unphysical momentum $k$ from the ground state contribution and
enables the extraction of the physical matrix elements $\langle
\eta_c (p)K(-q)|{\cal {\widetilde{O}}}(0)|B(p-q)\rangle$
 and $\langle \chi_{c0} (p)K(-q)|{\cal {\widetilde{O}}}(0)|B(p-q)\rangle$ due to the
simultaneous conditions, $P^2 = m_B^2$ and $(p-q)^2 = m_B^2$. The
light-cone distribution amplitudes $\phi_{3K}$, $
\tilde{\phi}_{\parallel}$, $\tilde{\phi}_{\perp}$,
$\phi_{\parallel}$ and $\phi_{\perp}$ are analytical  functions, as
well as only numerical values are concerned, we can analytically
continue $P^2$ into the timelike region, $P^2=m^2_B$, if the
unphysical negative momentum fractions of the partons (quarks and gluons) are  taken
for granted,  we can  then take the integral
$\int^{\alpha}_{\alpha_c}
 d\beta
G(s,x,\alpha,\beta)$ directly, the numerical results will not make
difference. In the channel $B\rightarrow \pi\pi$, the duality
region is about $s_{\pi}=0.7 GeV^2 \ll |P^2|\sim m_b^2$, the
quantity $s/(s-P^2)=-s/(P^2(1-s/P^2))$ is small, and the expansion
in terms of Taylor series of $s/P^2$ converges quickly, the
results will not  depend  significantly on the end point values
of the light-cone distribution amplitudes. As for the channels
$B\rightarrow  \eta_c K,\chi_{c0}K$, the duality regions are about
$s_{\eta_c}=11 GeV^2$ and $s_{ \chi_{c0}}=13 GeV^2$, the values of
$s/|P^2|$ are about $44\%$ and $52\%$, respectively. The expansion
in terms of Taylor series of $\beta$  for the light-cone
distribution amplitudes converges more slowly and more terms
should be taken into account, for example, $|\alpha_c| \leq 36\%$
in the $B\rightarrow \eta_c K$ channel and $|\alpha_c| \leq 55\%$ in the $B\rightarrow \chi_{c0}K$ channel;
the main drawback of this approach is the significant dependence on the
end point values of those light-cone distribution amplitudes.
\section{Numerical results }

Next, we choose the input parameters entering the light-cone sum
rules before giving numerical predictions on the  nonfactorizable
soft contributions.

The parameters enter the decays $B\longrightarrow \eta_c K,
\chi_{c0}K$ are taken as $m_B =  5.28$ GeV , $f_B = 180 \pm 30 $
MeV, $m_b = 4.7 \pm 0.1$ GeV, $s_{B} = 35 \pm 2\, {\rm GeV}^2$,
$m_{\eta_c} = 3.0$ GeV, $f_{\eta_c} = 0.35$ GeV , $m_c = 1.25 \pm
0.05$, $s_{\eta_c} = 11 \pm 1\, {\rm GeV}^2$,
 $m_{\chi_{c0}} = 3.41$ GeV, $f_{\chi_{c0}} = 0.36$ GeV,
$s_{\chi_{c0}} = 13 \pm 1\, {\rm GeV}^2$,
 and  $f_K = 0.16$ GeV \cite{Reinders}.  The value of decay constant $f_{\chi_{c0}} $ is
taken from Ref. \cite{Novikov} and new estimation based on the
nonperturbative approach of coupled Schwinger-Dyson equation and
Bethe-Salpeter equation is in preparation. For the coefficients of
the twist-3 and twist-4 kaon meson light-cone distribution
amplitudes, we can make an approximation $f_{3 \pi} \simeq f_{3
K}$, $\delta^2_K \simeq \delta_{\pi}^2$ and take $f_{3 K} =
0.0026$ GeV, $\delta^2(\mu_b) = 0.17$ GeV, where $\mu_b =
\sqrt{m_B^2 - m_b^2} \sim 2.4$ GeV \cite{Kho1,CZ,BF}.

Here we will take a short digression to discuss the duality
regions. In the $B\longrightarrow \eta_c K$ channel, as the
axial-vector current $J_{\mu}=\bar{c}\gamma_{\mu}\gamma_5c$ in
stead of pseudoscalar current $J_{5}=\bar{c} i\gamma_5c$ is chosen
to interpolate the $\eta_c $ meson, we must be careful in choosing
the duality region to avoid possible pollutions from the
$\eta_c(2s)$ and $\chi_{c1}$ mesons with the same quantum numbers
as the interpolating current. The masses of those two mesons are
about $m_{\eta_c(2s)}=3.6GeV$, $m_{\chi_{c1}}=3.5GeV$, and the
widths of those mesons are narrow, we can  choose the duality
region to be $s_{\eta_c} = 11 \pm 1\, {\rm GeV}^2$. In the
$B\longrightarrow  \chi_{c0} K$ channel, due to narrow width of
the $\chi_{c0}$ meson, we can choose the duality region to be
$s_{\chi_{c0}}=13\pm 1GeV^2$ to avoid possible pollutions from the
excited and continuum states. Furthermore, larger $s$ means larger
$|\alpha_c|$, more Taylor series of $\beta$ have to be re-summated, heavy
dependence on the end point values of the light-cone distribution
amplitudes. In those two channels, the variation of the duality thresholds $s_{\eta_c}$ and $s_{\chi_{c0}}$ can lead
to large uncertainties.

The parameters $n$ and $M^2$ must be carefully chosen to guarantee
the excited and continuum states to be  suppressed and to obtain a
reliable perturbative QCD calculation. The stable region for the
Borel parameter $M^2$ is found in the interval $M^2 = 10 \pm 2\,
{\rm GeV}^2$  which is known from the  $B$ channel QCD sum rules
\cite{Kho1}. In the charmonium channels, we usually perform n-th
derivative and take n-th moment sum rules to satisfy the stability
criteria  \cite{Reinders}.

For the decay $B\rightarrow \eta_c K$, the calculation is rather
stable on the range  $n = 3 - 7$. $Q_0^2$ is parameterized by
$Q_0^2 = 4 m_c^2 \xi$, where $\xi$ is usually allowed to take
values from 0 to 1 while the best interval is $\xi=0.3-1$.   In the following, we write
down the numerical value for the nonfactorizable soft matrix
element,
\begin{eqnarray}
 \langle \eta_c (p)K(q)|{\cal {\widetilde{O}}}(0)|B(p+q)\rangle= 0.035 \pm 0.010 GeV^3\, ,
\end{eqnarray}
the largest uncertainties come from the variations of the mass of
the c quark.

Taking into account the next-to-leading order Wilson coefficients calculated in the
naive dimensional regularization scheme \cite{Buras}
for $\mu = \overline{m_b}(m_b) = 4.40 {\rm \,
GeV}$ and $\Lambda_{\overline{\rm MS}}^{(5)} = 225\, {\rm MeV}$,
\begin{eqnarray}
\qquad C_1(\overline{m_b}(m_b)) = 1.082 \, , \qquad C_2(\overline{m_b}(m_b)) = -0.185 \, ,\qquad C_3(\overline{m_b}(m_b)) = 0.014 \, , \nonumber\\
\qquad C_4(\overline{m_b}(m_b)) = -0.035 \, ,\qquad C_5(\overline{m_b}(m_b)) = 0.009 \, ,\qquad C_6(\overline{m_b}(m_b)) = -0.041 \, ,
\end{eqnarray}
here we have neglected the Wilson coefficients $C_7,C_8,C_9,C_{10}$ in numerical
calculation due to their small values,  finally  we obtain the numerical relation between  the
contributions from the factorizable and nonfactorizable matrix
elements,
\begin{eqnarray}
\frac{\left\{2 V_{c b} V_{c s}^*C_1 (\mu)-2V_{t b} V_{t s}^*\left[
C_4 (\mu)+C_6 (\mu)  \right] \right\} \langle \eta_c (p)K(q)|{\cal
{\widetilde{O}}}(0)|B(p+q)\rangle} {\left\{ V_{c b} V_{c
s}^*\left[C_2(\mu)+\frac{C_1 (\mu)}{3}\right]-V_{t b} V_{t
s}^*\left[ C_3 (\mu)-C_5 (\mu) +\frac{C_4 (\mu)-C_6 (\mu)}{3}
\right]\right\}f_{\eta_c}m^2_B F^0(m^2_{\eta_c})}=0.11 \pm 0.04 \, .
\end{eqnarray}

From  above expressions, we can see that the nonfactorizable soft
contributions are considerable and they must be included in
analyzing the branching fraction. A rough estimation shows that the
theoretical branching fraction will increase to about $1.15-1.30$
times as the naive factorization result. Although there are still
large mismatches between the theoretical and experimental values,
we can say that the theoretical prediction is considerably
improved. The consistent and complete  QCD LCSR analysis should
include all the contributions from $O(\alpha_s)$  corrections,
however, the calculation is cumbersome and we prefer another
article.

 For the decay $B\rightarrow \chi_{c0}K$, the calculation is
rather stable on the range  $n = 5 - 10$, the parameter $\xi$ is
usually allowed to take values larger than 1 while the best
interval is $\xi=1.2-2.0$.
The nonfactorizable soft hadronic matrix element in the
$B\rightarrow \chi_{c0}K$ decay is
\begin{eqnarray}
 \langle \chi_{c0}
 (p)K(q)|{\cal {\widetilde{O}}}(0)|B(p+q)\rangle= 0.22 \pm 0.08 GeV^3 \, ,
\end{eqnarray}
and the corresponding branching fraction is $(1.0\pm 0.6) \times
10^{-4}$  which is  smaller than the experimental data $(6.0\pm
2.1) \times 10^{-4}$ \cite{Abe}. From Eq.(48) and Eq.(51), we can
see that value of the nonfactorizable soft hadronic matrix
element $\langle \eta_c (p)K(q)|{\cal
{\widetilde{O}}}(0)|B(p+q)\rangle$
 is about $15\%$ of the $\langle \chi_{c0} (p)K(q)|{\cal {\widetilde{O}}}(0)|B(p+q)\rangle$
 , and contributes to the branching fractions with magnitude about $10^{-5}$ .
Take into account for the contributions from the factorizable
hadronic matrix elements, the theoretical predicated branching
fraction  is about $10\%$ of the corresponding experimental data
for the decay $B\rightarrow \eta_c K$.
In this article, we take into account only the nonfactorizable soft
 contributions of twist-3 and twist-4 , while
higher twist contributions are neglected.  Although there are large uncertainties due to
 discarding the higher twist contributions and varying the duality thresholds, we can estimate the order of
magnitude of the  nonfactorizable soft effects at least.

 \section{conclusion}
In this article, we have analyzed the contributions from the
nonfactorizable soft hadronic matrix elements  $\langle \eta_c
(p)K(q)|{\cal {\widetilde{O}}}(0)|B(p+q)\rangle$
 and $\langle \chi_{c0} (p)K(q)|{\cal {\widetilde{O}}}(0)|B(p+q)\rangle$
   to the  decays
$B\rightarrow \eta_c K, \chi_{c0} K$ with  the effective weak
Hamiltonian. As the QCD-improved  factorization approach breaks
down in the B to charmonia decays, the contributions from the soft
gluon exchanges  will signalize  themselves. Our numerical results
show that their contributions are considerable and should  not be
neglected for the decay $B\rightarrow \eta_c K$. Although there
are still large mismatches  between the theoretical and
experimental values, the theoretical predicted branching fraction is considerably
improved (about $(15-30)\%$). As for the decay  $B\rightarrow
\chi_{c0} K$, there are no factorizable contributions, the
nonperturbative  contributions from the nonfactorizable soft
matrix elements are about five times smaller than  the
experimental data. The consistent  and complete  QCD LCSR analysis
should include all the contributions from $O(\alpha_s)$
corrections and higher twist contributions.

\section*{Acknowledgment}

The author (Z.G.Wang) would like to thank National
Postdoctoral Foundation for financial support.

\end{document}